\documentclass[10pt,twoside,BCOR7mm,DIV15,headinclude,footexclude,cleardoubleempty,idxtotoc]{scrartcl}

\usepackage[english]{babel}
\usepackage{graphicx}
\usepackage{ifthen}
\usepackage{scrpage2}
\usepackage{hyperref}

\makeatletter
\renewcommand{\@biblabel}[1]{}
\renewcommand{\@cite}[2]{%
{#1\ifthenelse{\boolean{@tempswa}}{,#2}{}}}
\makeatother

\hypersetup{breaklinks=true
,colorlinks=true,linkcolor=black,urlcolor=blue
,citecolor=black}

\ofoot{\thepage}
\ifoot{}

\setheadsepline{1pt}

\setkomafont{pagehead}{\normalfont\sffamily}
\setkomafont{pagenumber}{\normalfont\rmfamily}

\usepackage{booktabs}
\usepackage{amsmath}
\usepackage{amssymb}
\usepackage{multicol}
\usepackage{float}

\makeatletter
\newcommand{\listofcontributions}{\@starttoc{con}}

\newcommand{\l@contribution} {\@dottedtocline{1}{1.5em}{2.3em}}
\makeatother

\newenvironment{contribution}{
\setcounter{section}{0}
\setcounter{figure}{0}
\setcounter{table}{0}
\begin{flushleft}
{\em Clumping in Hot Star Winds \\
W.-R.\ Hamann, A.\ Feldmeier \& L.\ Oskinova, eds.\\
Potsdam: Univ.-Verl., 2007 \\
URN: http://nbn-resolving.de/urn:nbn:de:kobv:517-opus-13981
} 
\end{flushleft}
}{
\newpage
\lehead{}
\rohead{}
}

\begin{document}

\setlength{\baselineskip}{2.5ex}

\begin{contribution}
\newcommand{\lsim}{\raisebox{-.4ex}{$\stackrel{<}{\scriptstyle \sim}$}}
\newcommand{\msim}{\raisebox{-.4ex}{$\stackrel{>}{\scriptstyle \sim}$}}
\lehead{L.M.\ Oskinova, W.-R.\ Hamann \& A.\ Feldmeier}
\rohead{X-raying clumped stellar winds}
\begin{center}
 {\LARGE \bf X-raying clumped stellar winds}\\
 \medskip
 {\it\bf L.M.\ Oskinova$^1$, W.-R.\ Hamann$^1$ \& A.\ Feldmeier$^1$} \\
 {\it $^1$Universit\"at Potsdam, Germany}\\
%
\begin{abstract}
{X-ray spectroscopy is a sensitive probe of stellar winds.  X-rays
originate from optically thin shock-heated plasma deep inside the wind
and propagate outwards throughout absorbing cool material.  Recent
analyses of the line ratios from He-like ions in the X-ray spectra of
O-stars highlighted problems with this general paradigm: the measured
line ratios of highest ions are consistent with the location of the
hottest X-ray emitting plasma very close to the base of the wind,
perhaps indicating the presence of a corona, while measurements from
lower ions conform with the wind-embedded shock model.  Generally, to
correctly model the emerging X-ray spectra, a detailed knowledge of
the cool wind opacities based on stellar atmosphere models is
prerequisite. A nearly grey stellar wind opacity for the X-rays is
deduced from the analyses of high-resolution X-ray spectra. This
indicates that the stellar winds are strongly clumped. Furthermore,
the nearly symmetric shape of X-ray emission line profiles can be
explained if the wind clumps are radially compressed.  In massive
binaries the orbital variations of X-ray emission allow to probe the
opacity of the stellar wind; results support the picture of strong
wind clumping. In high-mass X-ray binaries, the stochastic X-ray
variability and the extend of the stellar-wind part photoionized by
X-rays provide further strong evidence that stellar winds consist of
dense clumps.}
\end{abstract}
\end{center}
%
\begin{multicols}{2}
\section{Introduction}

All OB-stars with spectral types earlier than B1.5 have X-ray
luminosities $L_{\rm X}$ exceeding $10^{30.5}$\,erg\,s$^{-1}$.
Remarkably, their X-ray luminosity roughly correlates with stellar
bolometric luminosities as $L_{\rm X}$$\approx$$10^{-7}L_{\rm bol}$
(Seward et al. \cite{sew79}, Bergh\"ofer et al. \cite{berg97}).  The
high-resolution X-ray spectra of two O-stars, $\zeta$\,Pup and
$\zeta$\,Ori, are shown in Figure\,1. The spectra are dominated by
strong emission lines that reflect stellar wind abundances, e.g.
nitrogen lines are more prominent in the spectrum of the more evolved
star $\zeta$\,Pup. X-ray spectra of single
non-magnetic massive stars can be explained as emitted by optically
thin plasma heated to  temperatures $kT_{\rm X}$$\lsim$$0.6$\,keV
(e.g. Wojdowski \& Schulz \cite{ws05}).

There is a number of hypotheses to explain how X-rays are generated in
stellar winds. Lucy \& White (\cite{lw80}) put forward a theory
where X-rays  originate from bow shocks around a
population of blobs (see also Cassinelli et al., these proceedings).
Cassinelli \& Swank (\cite{csw83}) proposed an analogy with coronal
stars where X-rays originate from magnetically confined regions at the
wind base. Owocki et al.\ (\cite{ocr88}) performed numerical
simulations of the evolution of instabilities in radiatively
driven stellar winds and predicted a development of reverse shocks
with postshock temperatures in the range $10^{6}-10^{7}$\,K (i.e. as
inferred from the observed X-ray spectra).  Feldmeier et al.\
(\cite{feld97}) developed a model explaining the X-rays as
originating from cooling zones behind shock fronts and were able to
reproduce the low-resolution {\em Rosat} spectra of $\zeta$\,Pup
and $\zeta$\,Ori.

In massive binaries a copious amount of X-ray emission is produced by
the collision of two powerful stellar winds. X-ray spectra of
colliding wind binaries reveal that the X-ray emitting plasma departs
from collisional equilibrium (e.g. Pollock et al. \cite{amtp05}).
Colliding wind systems tend to have somewhat higher X-ray luminosities
and their X-ray spectra can be ``harder'' compared to the single stars.
%
\begin{figure}[H]
\begin{center}
\includegraphics[width=0.7\columnwidth]{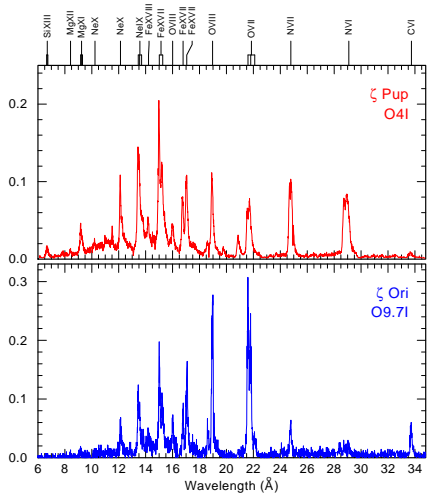}
\caption{High-resolution X-ray spectra of $\zeta$\,Pup 
{\it (top panel)} and $\zeta$\,Ori {\it (bottom panel)} obtained 
with {\em XMM-Newton}. Prominent emission lines are identified.}
\label{fig:der}
\end{center}
\end{figure}
%
Another type of X-ray emitting plasma is found in high-mass X-ray
binary systems (HMXBs) that consist of an OB supergiant and a
degenerate companion, usually a neutron star (NS) deeply embedded in
the stellar wind of the primary.  Stellar wind accretion onto a NS
powers strong X-ray emission that has a characteristic power-law
spectrum, while the surrounding stellar wind is photoionized by this
X-ray emission.

Independent on the formation mechanism, the X-rays registered by the
observer are attenuated in the stellar wind and interstellar medium.
The transport of X-rays in stellar winds can often be treated as pure
absorption (Baum et al.\ \cite{baum92}, Hillier et al.\ \cite{hil93}).
This leads to a great simplification of the radiative transfer problem
and allows modeling even in the case when the absorbing medium is
inhomogeneous.

\section{X-ray emission of single stars as a probe of clumped stellar 
wind}

\subsection{Line ratios in spectra of He-like ions}

One of the key spectral diagnostics is provided by the ratio of line
fluxes from He-like ions. These ions show characteristic '{\em fir}
triplets' of a forbidden (f), an intercombination (i) and a resonance
(r) line. In OB-type stars, the $\cal{R}\equiv$$f/i$ line ratio is
sensitive to the local mean intensity of the UV radiative field,
$J_{\nu}$. The UV photons with wavelength $\lambda_{{\rm f}\rightarrow
{\rm i}}$ excite the metastable $^3$S level to the $^3$P level, so
that ${\cal R}$$\propto$$(1+ \phi(J_{\nu}))^{-1}$, where
$\phi(J_{\nu})$ is the photoexcitation rate (constants and collisional
terms are omitted).  Neglecting the limb darkening, the average
intensity can be substituted by the diluted photospheric flux,
$W(r)H_{\nu}$. {\em If} the photospheric flux $H_{\nu}$ is known, the
dilution factor $W$ can be inferred from ${\cal R}$. This constrains
the radius $R_{\rm fir}$ where the X-ray emitting plasma is located.

For wavelengths in the observable part of the UV the flux can be, in
principle, directly inferred from observations.  Unfortunately,
especially for giant and supergiant stars, the photospheric fluxes are
often contaminated by wind lines. E.g.\ O\,{\sc vi} resonance doublet
coincides with the $\lambda_{{\rm f} \rightarrow {\rm i}}$ transition
at 1033\,\AA\ for the Mg\,{\sc xi} line.  The O\,{\sc vi} doublet
itself can only be reproduced by stellar atmosphere models if an X-ray
field causing Auger ionization is included (e.g.\ Oskinova et al.\
\cite{osk06}). This example highlights the need of a full radiative
transfer treatment for the correct interpretation of the {\em fir}
line ratios -- a task that has not been accomplished yet.

Commonly, the fluxes $H_{\nu}$ at the wavelengths of interest are
obtained from stellar atmosphere models.  Leutenegger et al.\
(\cite{leu06}) analyzed the lines of He-like ions in the spectra of
four O-type giants and supergiants. The parameters of the O-stars were
taken from Lamers \& Leitherer (\cite{ll93}) and used to select a
stellar atmosphere model from the TLUSTY static plane-parallel model
grid. Leutenegger et al.\ find that the minimum radius of X-ray
formation is typically in the range of $1.25$$<R_{\rm fir}/R_*$$<1.67$.
%
\begin{figure}[H]
\begin{center}
\includegraphics[width=0.75\columnwidth]{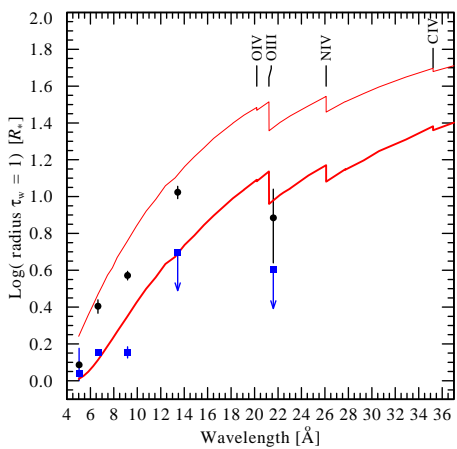}
\caption{Logarithm of the radius were the radial optical depth of the 
wind becomes unity in  $\zeta$\,Pup according to PoWR
stellar atmosphere models. Thick line: model with 
$\dot{M}$$=$$2.5$$\times$$10^{-6}\,M_\odot/{\rm yr}$;  thin line:  
$\dot{M}$$=$$8.7\times$$10^{-6}\,M_\odot/{\rm yr}$. The radii of the
X-ray emission derived from different He-like ions are indicated by
dots (from WC07) and squares (Leutenegger et al.\ 2006).}
\label{fig:fir}
\end{center}
\end{figure}
%

Waldron \& Cassinelli (\cite{wc07}, WC07) analyzed {\em fir} line
ratios for 17 OB stars, using plane-parallel Kurucz models, and found,
in agreement with their earlier results, that high-$Z$ ions are
predominantly located closer to the photosphere. They highlighted a
{\em ``near-star high-ion'' problem}. This problem can be illustrated
by the S\,{\sc xv} line in $\zeta$\,Puppis. WC07 constrained $R_{\rm
fir}$ for this line to $<1.22\,R_*$. This is in agreement with the
results of Leutenegger et al.\ (\cite{leu06}) who estimated the
minimum value of $R_{\rm fir}$ as $1.1^{+0.4}_{-0.1}\,R_*$. The
emissivity of S\,{\sc xv} has its maximum at $T\approx$$16$\,MK.  The
presence of plasma with such high temperatures at $\approx$$1.1\,R_*$,
i.e.  close to the photosphere, is contrary to the expectations from
the shocked-wind model, but supports the base-corona model.

WC07 notice a correlation between the radius where the ambient cool wind
becomes transparent for the X-rays $R(\tau_{\rm wind})$$=$$1$, and the
predominant location of He-like ion X-ray emission. This is
illustrated in Fig.\,2.  We use a spherically-symmetric PoWR
atmosphere model of $\zeta$\,Pup to calculate $R(\tau_{\rm
wind})$$=$$1$.  The calculations are performed for two different
$\dot{M}$ evaluated under the assumption of unclumped wind (Repolust
et al.\ \cite{rep04}) and allowing for macroclumping (Oskinova et al.\
\cite{osk07}).  The measurements of $R_{\rm fir}$ by WC07 and
Leutenegger et al.\ (\cite{leu06}) are also shown. As can be seen in
Fig.\,2, there is a good agreement between the measurements by WC07 of
the {\em fir}-inferred radii and $R(\tau_{\rm wind})$$=$$1$.  However,
there is an apparent disagreement with the results of Leutenegger et
al.\ (\cite{leu06}).  This can be due to a difference in measuring
${\cal R}$ and/or different approaches to the analysis (see WC07).

Because we preferentially see X-rays that are generated close to the
$\tau_{\rm wind}=1$ surface (e.g. Ignace et al.\, \cite{ig00}), the
results of Leutenegger et al.\ imply a reduced mass-loss rate, while
the WC07 results are consistent with the $\dot{M}$ estimates of Puls
et al.\ (\cite{puls06}) and Oskinova et al.\ (\cite{osk07}).


\subsection{X-ray emission line profiles}

%
\begin{figure}[H]
\begin{center}
\includegraphics[width=0.7\columnwidth]{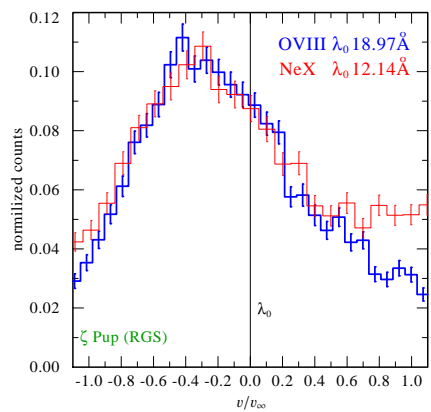}
\caption{O\,{\sc viii} (thick blue) and Ne\,{\sc x} (thin red) lines
in the spectrum of $\zeta$ Pup plotted versus wavelength in 
velocity units. The red-ward shoulder in Ne\,{\sc x} line is due an
unresolved Fe emission feature. }
\label{fig:fig3}
\end{center}
\end{figure}
%

The analysis of {\em fir}-lines suggests that lines from lower $Z$
ions are formed further out in the wind. Therefore, assuming
that the wind velocity monotonically increasing  with the radius,
lines from ions with lower {\em Z} should be broader than from ions
with higher {\em Z}. Contradictory to this expectation, the observed
widths of all X-ray lines are similar (e.g Kramer et al.\ \cite{kra03};
Pollock \cite{pol07}).  To illustrate this, the lines of Ne\,{\sc x}
and O\,{\sc viii} in spectrum of $\zeta$\,Pup are shown in Fig.\,3
(see also lines of C\,{\sc vi} and Ne\,{\sc x} in $\zeta$\,Ori
spectrum shown in Pollock \cite{pol07}). Thus, there is a discrepancy
between the location of the emission region inferred from the {\em fir}
line ratios and from the fitting of the line profiles.

Radiative transfer in homogeneous wind predicts that the redshifted
emission from the back side of the wind is more suppressed by the
continuum opacity along the line of sight, than the blueshifted
emission from the front hemisphere. This should lead to skewed, almost
triangular-shaped profiles of the emission lines (MacFarlane
et\,al. 1991; Ignace \cite{ri01}; Owocki \& Cohen \cite{oc01}). Because 
the atomic opacity $\kappa_\nu$ increases at longer wavelengths, the 
lines of ions with lower $Z$ are expected to be more skewed compare to 
the higher $Z$ ions. However, the majority of observed stellar X-ray lines 
are only slightly skewed and the shifts are similar for lines of 
all ions (WC07).

There are a number of different hypotheses to explain the
puzzling properties of observed line profiles. Pollock
(\cite{pol07}) proposes that X-rays originate in collisionless
shocks. Two-component wind expansion is invoked by Mullan \& Waldron
(\cite{mw06}). Ignace \& Galley (\cite{ri02}) show that if the X-ray
emitting plasma is optically thick this would result in more symmetric
line profiles (see also Leutenegger et\,al., these proceedings).  Waldron
\& Cassinelli (\cite{wc01}) suggest that nearly symmetric line
profiles can be explained by low opacity of the stellar wind.

Waldron \& Cassinelli (\cite{wc01}) note that the wind opacity could
be reduced if a large fraction of the wind were photoionized by
X-rays. However, they observe that this is not compatible with the
presence of low-ionisation species, e.g.\ C\,{\sc iv} in the UV
spectra.  Our recent PoWR models of O star winds confirm this
suggestion: there is no significant change in the ionisation structure
of the wind when a realistic X-ray intensity is included.

Waldron \& Cassinelli (2001) further suggest that if the adopted
mass-loss rate is reduced this would lead to a better agreement
between modeled and observed line shapes.  This suggestion is
supported by the modeling of line profiles emerging from a homogeneous
wind by Kramer et al.\ (2003) and Cohen et al.\ (2006). They infer the
wind optical depth by fitting a model with four free parameters to the
lines observed in the X-ray spectra of $\zeta$\,Pup and
$\zeta$\,Ori. The optical depth in the smooth wind with a monotonic
velocity law scales with a parameter $\tau_*= \kappa_{\nu}\dot{M}
(v_\infty R_*)^{-1}$ (Owocki \& Cohen \cite{oc01}).  Inferring
$\tau_*$ from the model fits and assuming that the wind atomic opacity
$\kappa_{\nu}$ is a constant, Kramer et al.\ (2003) and Cohen et al.\
(2006) conclude that X-ray lines can be fitted only if $\dot{M}$ for
$\zeta$\,Pup and $\zeta$\,Ori are strongly reduced (see also Cohen,
these proceedings).  More realistic mass-absorption coefficients
$\kappa_\nu$ are presented in Oskinova et al.\ (2006). We use the
best-fit values of $\tau_*$ obtained for different lines from Table\,1
in Kramer et al.\ (2003) to estimate the mass-loss in $\zeta$\,Pup,
however there is large disagreement  when different lines are
considered.

The smooth-wind line profile formalism can be easily adapted for an
inhomogeneous wind, where {\em all} clumps are optically thin at {\em
any} given wavelength (so-called microclumping approximation). The
empirical mass-loss rates based on $\rho^2$ diagnostics are lower when
microclumping is adopted, compared to models that assume a smooth
wind. The mass-loss rates obtained from the fitting of X-ray lines by
Kramer et al.\ (2003) can be reconciled with the radio and H$\alpha$
measurements only if clumping filling factors are very small (see Puls
et al.\ \cite{puls06}). Overall, the use of microclumping
approximation, at least in dense winds, is questionable (see Hamann,
these proceedings).
 
Waldron \& Cassinelli (2001) also note that a {\em clumped wind} can
be effectively optically thin even in stars with large $\dot{M}$.
This suggestion was exploited in Feldmeier et al.\ (\cite{feld03}) by
waiving the microclumping approximation.

Assume that the flow of clumps that constitutes the wind obeys the
equation of continuity. The number of clumps per unit volume is
$n(r)$$\equiv$$n_0v(r)^{-1}r^{-2}$, where $n_0$ is a
constant. The effective opacity of clumped wind, $\kappa_{\rm
eff}=n(r) \sigma_{\rm clump} {\cal P}$, is the product of the clump
number density $n(r)$, the clump cross-section
$\sigma_{\rm clump}$, and the probability of an X-ray
photon being absorbed when it encounters a clump, ${\cal
P}=1-\exp{(-\tau^{\rm clump}_\nu)}$ (Feldmeier et al.\ \cite{feld03}).

The optical depth across an average clump at the distance $r$ is
$\tau^{\rm clump}_\nu=\tau_* R_* v_\infty r^{-2}n_0^{-1}$.  When
$\tau^{\rm clump}_\nu \gg 1$ the effective opacity $\kappa_{\rm eff}$
does not depend on the atomic opacity $\kappa_\nu$ (because $\cal{P}
\rightarrow 1$). Therefore, the wind effective opacity becomes grey.

To calculate emission line profiles, we assume for simplicity that the
emission with emissivity $\eta$ originates between some specific radii
$r_1$ and $r_2$ in the wind. The formal integral for the line profile
reads
$ 
F_\mu\sim\int_{r_1}^{r_2}
\eta (r)r^2 {\rm e}^{-\tau_{\rm w}} {\rm d}r, 
$ 
where $\mu$ is the direction cosine.

Evaluating the wind optical depth as an integral over effective opacity
along the line-of-sight $z$ gives: 
\begin{equation} 
\tau_{\rm w}=n_0\int_{z_\nu}^\infty \frac{\sigma}{v(r)r^2} 
(1-{\rm e}^{-\tau^{\rm clump}_\nu}) {\rm d}z. 
\label{eq:tol} 
\end{equation} 
To understand how
the clump geometry alters the line profile shape, let us consider two
extreme cases and compare isotropic clumps (balls) and anisotropic clumps
in the form of infinitesimally thin shell-fragments (pancakes) oriented
perpendicular to the radial direction, so-called a ``Venetian blind''
model. Assuming that the clumps keep constant solid angle as they
propagate outwards, in the direction $z$ a ball has the cross-section
$\sigma$$\propto$$r^2$, while a pancake has $\sigma$$\propto$$|\mu|r^2$.  
Recalling that $dz=dr/\mu$ and inserting the above expressions for the
cross-sections in the Eq.\,(\ref{eq:tol}), it is immediately clear that in
the case of pancakes the integral Eq.\,(\ref{eq:tol}) transforms into an
integral over $r$, while it stays an integral over $z$ in case of the
balls. The consequence is that pancakes yield nearly symmetric emission
line profiles, similarly to the observed (Oskinova et al.\ \cite{osk06}).

Owocki \& Cohen (2006) applied a porosity length formalism (Owocki et
al.\ \cite{ow04}) with $\kappa_{\rm eff}=h(r)^{-1}(1-\exp(-\tau^{\rm
clump}_\nu))$, where $h(r)$ is a parameter. They studied a case of
isotropic opacity, and concluded that large porosity lengths
$h(r)$$\approx$$1R_*$ are required to reproduce the observed X-ray
line profiles. However, Oskinova et al.\ (2006) computed synthetic
X-ray lines for anisotropic opacity. Stellar wind parameters, i.e.
$\kappa_\nu$, $\dot{M}$, and $v(r)$ were adopted from most recent
stellar atmosphere models and it was assumed that the average
separation between clumps at distance $r$ is $R_*v(r)/v_\infty$, as
compatible with predictions of hydrodynamic simulations (e.g Feldmeier
et al.\ 1997).  Synthetic lines were computed {\em without allowing
any free parameter} and compared with the observed lines. The
remarkable similarity between synthetic and observed lines provides an
evidence that the wind clumps are not optically thin, and are
compressed in radial direction.


\section{X-ray emission of binaries as a probe of clumped 
stellar wind}

The collision of stellar winds in massive binary systems (CWB) results
in a bow shock that concaves around the star with the weaker wind. In
close binaries the shocked material cools radiatively and hence the
intrinsic X-ray luminosity is $L_{\rm X}$$\propto$$\dot{M}v^2$
(Pittard \& Stevens
\cite{pit02}). In the wide binaries with sufficiently large separation 
$d$, the shocked material cools adiabatically resulting in $L_{\rm
X}$$\propto$$\dot{M}^2d^{-1}v^{-1.5}$ (Stevens et al.\
\cite{ste92}).

Hydrodynamic studies of colliding {\em clumped} winds were conducted
by Walder \& Folini (2002) and Pittard (2007).  Walder \& Folini
considered WR\,140 at periastron. They showed that the collision of
clumped winds in a radiatively cooling system leads to the
fragmentation of the colliding wind region (CWR). Carbon enriched
dense WR-wind clumps with dimensions exceeding $10^{11}$\,cm can
effectively cool and serve as the seeds for the formation of the
dust. Pittard (2007) considered WR\,140 at apastron when the CWR is
likely to cool adiabatically. It was shown that in this case the CWR
becomes highly turbulent. This leads to a rapid destruction of the
clumps in the CWR. Therefore in adiabatic CWB, wind clumping does not
affect the intrinsic X-ray luminosity. Therefore such systems may
provide a diagnostic of mass-loss that is independent on
clumping.

The X-rays generated in the bow shock propagate along the line of sight
through the stellar wind of the foremost star, resulting in X-ray
eclipses due to the orbital motion.  The duration and the depth of the
eclipses depends on the orbital parameters, the wind geometry, and its
opacity.  When the intrinsic X-ray luminosity from the bow shock is
known, the stellar mass-loss rate can be probed by inferring the
absorbing column from the X-ray spectrum.

Schild et al.\ (\cite{sch04}) derived the mass-loss rate of the WC8
component of $\gamma$\,Vel in this way. The obtained $\dot{M}$ is a
factor of four smaller than what is derived from spectral analysis
with homogeneous wind models. This was interpreted in terms of WR wind
clumping, with a volume filling factor of $f_{\rm V}\approx 0.06$. For
$\eta$\,Car, Pittard \& Corcoran (2002) derived a mass-loss rate from
an analysis of the X-ray spectrum that is lower by a factor of few
compared to the conventionally adopted value, thus indicating wind
clumping.  Similarly, Pollock et al.\ (\cite{amtp05}) derived smaller
than expected wind absorption from their analysis of the X-ray
spectrum of WR140.

Supergiant HMXBs consist of a supergiant OB star and (usually) a
neutron star (NS), that orbits deep inside the stellar wind. An X-ray
emission with a power-law spectrum results from the accretion of the
stellar wind onto the NS. These X-rays photoionize the surrounding
stellar wind. The wind X-ray spectrum shows a large variety of
emission features, including a number of fluorescent lines.  Sako et
al.\ (2003) reviewed the spectroscopic results obtained with X-ray
observatories for several wind-fed HMXBs. They concluded that the
observed spectra can be explained only as originating in a wind where
cool dense clumps are embedded in rarefied photoionized gas. Sako et
al.\ (1999) constrained the volume filling factor of clumps in the
wind of Vela X-1 as $f_{\rm V}\approx$$0.04$. Similar conclusions were
reached by Van der Meer et al.\ (\cite{vdm05}). They studied the X-ray
light curve of 4U 1700-37 and found that the feeding of the NS by
stellar wind clumps explains the observed stochastic variability. The
clump separation at the distance $2R_*$ was constrained to $0.4R_*$.
Recently, the advances in the $\gamma$-ray and X-ray observations lead
to the discovery of new HMXBs that are highly absorbed and display
fast stochastic variability. This spectral and temporal variability is
consistent with strong wind clumping. The inferred
clumping parameters are similar to those obtained from the analysis of
X-ray emission in single stars and in CWBs (see Romero et al. and
Walter et al., these proceedings).
   
\section{Conclusions}
The spectral and temporal properties of the X-ray emission from single
stars, colliding wind binaries and high-mass X-ray binaries provide
consistent evidence of wind clumping. The multitude of data requires
strongly clumped winds where clumps are separated by a few tenths of
the stellar radius in the wind acceleration zone. The approximation
that clumps are optically thin at {\em all} wavelengths in the X-ray
band is not justified and cannot be applied universally. Clumped
stellar winds have a reduced opacity for X-rays, compared to a
homogeneous wind, due to a reduction of empirically determined
mass-loss rates and to the effect of wind porosity.
   

\end{multicols}




\end{contribution}


\end{document}